\documentclass[aps,prl,twocolumn,groupedaddress]{revtex4}
\usepackage{graphicx}

\begin{document}

%\preprint{}

\title {Mixed magnetic phases in (Ga,Mn)As epilayers}



\author{K. Hamaya, T. Taniyama$^{1}$, Y. Kitamoto, and Y. Yamazaki}
\affiliation{%
Department of Innovative and Engineered Materials, Tokyo Institute of
Technology,\\ 4259 Nagatsuta, Midori-ku, Yokohama 226-8502, Japan.\\
$^{1}$Materials and Structures Laboratory, Tokyo Institute of Technology,\\ 4259 Nagatsuta, Midori-ku, Yokohama 226-8503, Japan.\\
}%



\date{\today}

\begin{abstract}
Two different ferromagnetic-paramagnetic transitions are detected in (Ga,Mn)As/GaAs(001) epilayers from ac susceptibility measurements: transition at a higher temperature results from (Ga,Mn)As cluster phases with [110] uniaxial anisotropy and that at a lower temperature is associated with a ferromagnetic (Ga,Mn)As matrix with $\left\langle 100 \right\rangle$ cubic anisotropy. A change in the magnetic easy axis from [100] to [110] with increasing temperature can be explained by the reduced contribution of $\left\langle 100 \right\rangle$ cubic anisotropy to the magnetic properties above the transition temperature of the (Ga,Mn)As matrix.

\end{abstract}
%insert suggested PACS numbers in braces on next line
%\keywords{}


\maketitle

%\section{Introduction}
%\section{INTRODUCTION}
%\subsubsection{}

Ferromagnetism in Mn-doped $p-$type semiconductors can theoretically be understood by the $p-d$ exchange interaction between hole carriers and doped Mn spins \cite{Ohno2,Dietl2,Abolfath,Keavney}, which can be manipulated by electric-field \cite{Ohno} or optical-hole generation \cite{Oiwa}.
Recent studies of magnetic anisotropy in (Ga,Mn)As/GaAs(001) epilayers, on the other hand, have shown a significant change in the magnetic anisotropy with increasing temperature \cite{Sawicki,Welp,Liu,Hamaya2}; uniaxial anisotropy along [110] ([110] uniaxial anisotropy) \cite{Tang,Hrabovsky,Sawicki,Welp,Liu,Hamaya2} becomes predominant with increasing temperature \cite{Sawicki,Welp,Liu,Hamaya2} as well as increasing hole concentration\cite{Hamaya4}. In spite of these intensive studies, the origin of [110] uniaxial anisotropy is still unclear within the current theoretical framework and an understanding of the change in the magnetic anisotropy with temperature is therefore of fundamental importance in revealing the physics of magnetic semiconductors.

The aim of this study is to give a comprehensive description of the magnetic anisotropy in (Ga,Mn)As epilayers. In this Letter, two different peaks in temperature-dependent ac susceptibility that are associated with ferromagnetic-paramagnetic transitions are clearly shown. The peak profile at a higher transition temperature has a considerable dependence on frequency, indicating a blocking process in magnetic clusters, while that at a lower transition temperature shows no frequency dependence. The results provide evidence that the magnetic transitions at the lower and higher temperatures are associated with magnetic phases with cubic anisotropy and [110] uniaxial anisotropies, respectively, and the crossover of the magnetic easy axis from $\left\langle 100 \right\rangle$ to [110]  can be interpreted by the ferromagnetic-paramagnetic transition of the magnetic phase with $\left\langle 100 \right\rangle$ cubic magnetocrystalline anisotropy accordingly.

%\section{EXPERIMENTS}
100 nm-thick (Ga,Mn)As epilayers were grown on top of a 400-nm-thick GaAs buffer layer grown at 590$^{\circ}$C on a semi-insulating GaAs (001) substrate using low-temperature molecular beam epitaxy (MBE) at 190 $-$ 235$^{\circ}$C under an As-rich growth condition. To increase the hole carrier concentration, some epilayers were subject to post-growth annealing in an N$_{2}$ atmosphere for  60 $-$ 240 min at 250$^{\circ}$C \cite{Hayashi}; hole concentration was controlled by changing the Mn content and/or low-temperature annealing \cite{Hamaya4}. Hole carrier concentrations $p$ ($=$ ionized Mn acceptor concentration) were measured with electrochemical capacitance-voltage (ECV) method at room temperature \cite{Moriya}.
In order to check a possible origin of magnetic anisotropy due to lattice strain, we characterized the crystal structures of all samples we grew using high resolution x-ray diffraction. The x-ray diffraction experiments revealed that lattice strain along [110] and [1$\overline{1}$0] were equivalent; the (Ga,Mn)As unit cell was not elongated asymmetrically toward either of the [110] and [1$\overline{1}$0] \cite{Hamaya4} so that we could rule out structural origins for the magnetic anisotropy. No precipitated phases were also detected in any of the samples. Magnetic properties were measured with a superconducting quantum interference device (SQUID) magnetometer and a physical property measurement system (PPMS).

%\section{RESULTS AND DISCUSSION}
Figure 1(a) shows the temperature-dependent magnetization ($M-T$ curves) of a typical (Ga,Mn)As epilayer with $p$ = 4.0 $\times$ 10$^{20}$ cm$^{-3}$ and a Curie temperature ($T_{c}$) of $\sim$ 62 K at various applied field orientations of [100], [110], and [1$\overline{1}$0] . A dc magnetic field of $H =$ 20 Oe was applied after the magnetization had almost saturated in a high magnetic field of $H =$ 70 kOe at 4 K.
In the low temperature regime ($T \lesssim$ 25 K), the magnetization is the largest for [100] of the three applied field orientations, indicating that the magnetic easy axis is parallel to [100] as indicated in recent reports \cite{Welp,Liu,Sawicki,Hamaya2}.
However, a crossover of the magnetic easy axis from [100] to [110] is seen at 25 $-$ 30 K, being similar to previous reports which show that the predominant magnetic anisotropy changes from $\left\langle 100 \right\rangle$ cubic magnetocrystalline anisotropy to [110] uniaxail anisotropy \cite{Welp,Liu,Hamaya2,Sawicki}.
\begin{figure}[t]
\includegraphics[width=8.5cm]{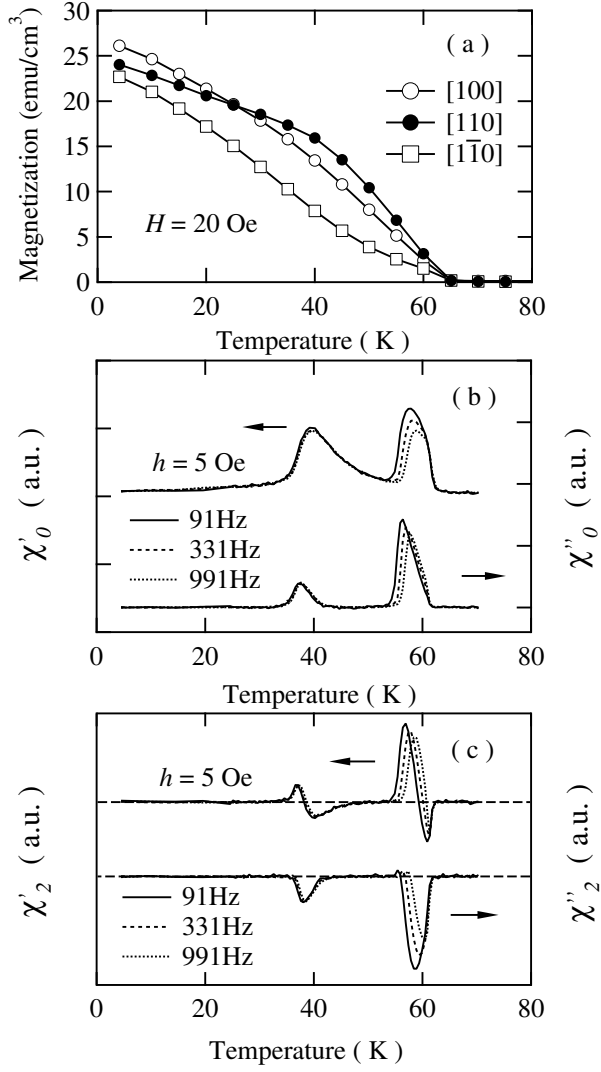}%%
\caption{(a) Temperature-dependent magnetization of a (Ga,Mn)As epilayer with a hole concentration of $p =$ 4 $\times$ 10$^{20}$ cm$^{-3}$ at various magnetic field orientations. The temperature dependence of (b) the linear and (c) the nonlinear susceptibilities as a function of frequency measured in $h =$ 5 Oe. }
\end{figure}

We measured the  temperature dependent ac susceptibility in an ac magnetic field $h$. In general, magnetization $m$ can be written as \cite{Bitoh}
\begin{equation}
{m = m_{0} + \chi_{0}h + \chi_{1}h^{2} + \chi_{2}h^{3} + ....} ,
\end{equation}
where m$_{0}$ is the spontaneous magnetization, $\chi_{0}$ is the linear susceptibility, and $\chi_{1}$, $\chi_{2}$,.. are the nonlinear susceptibilities. When the magnetic field is expressed as $h =$ $h_{0} sin$ $\omega$$t$, an voltage $E$ is induced in a pick-up coil as
\begin{equation}
E = A\{\chi^{t}_{0}h_{0}cos\omega t  + \chi^{t}_{1}h^{2}_{0}sin2\omega t - \frac{3}{4}\chi^{t}_{2}h^{3}_{0}cos3\omega t - ... \},
\end{equation}
where A is a numerical factor depending on the dimensions of the coil and the filling factor of a sample, and $\chi$$^{t}_{0}$ $=$ $\chi_{0}$ $+$ $\frac{3}{4}$$\chi_{2}$$h^{2}_{0}$ + $\frac{5}{8}$$\chi_{4}$$h^{4}_{0}$ $+$..., $\chi$$^{t}_{1}h_{0}$ $=$ $\chi_{1}$$h_{0}$ $+$ $\chi_{3}$$h^{3}_{0}$ $+$ $\frac{15}{16}$$\chi_{5}$$h^{5}_{0}$ $+$..., $\frac{3}{4}$$\chi$$^{t}_{2}h^{2}_{0}$ $=$ $\frac{3}{4}$$\chi_{2}$$h^{2}_{0}$ $+$ $\frac{15}{16}$$\chi_{4}$$h^{4}_{0}$ $+$...,.
If $h_{0}$ is small, the higher order ac signals at $\omega$ and 3$\omega$ are directly related to the linear and nonlinear susceptibilities, $\chi$$_{0}$ and $\chi$$_{2}$.

Figures 1(b) and (c) show the temperature dependence of both in-phase $\chi'$ and out-of-phase $\chi''$ components of linear ($\chi$$_{0}$) and nonlinear ($\chi$$_{2}$) susceptibilities of the same sample used in Fig. 1(a). The measurements were made with an ac magnetic field amplitude of $h =$ 5 Oe at a frequency $\omega$ of 91, 331, and 991 Hz, and the ac magnetic field was applied parallel to $\left\langle 100 \right\rangle$. Interestingly, two peaks are seen in the temperature-dependent $\chi'_{0}$ and $\chi''_{0}$ at all the frequencies we examined. The peak near $T_{c}$ ($\sim$ 60 K) is relatively sharp compared with that at $\sim$ 37 K.
Also, $\chi_{2}'$ and $\chi_{2}''$ have peaks and dips at around the same temperature. When single characteristic time $\tau$ describes the magnetic relaxation, dynamic susceptibility can be expressed as the Debye-type relaxation: $\chi(\omega)' =[1/(1+(\omega\tau)^2)]\chi_{0}$ and $\chi(\omega)'' =[1/(1+(\omega\tau)^2)](\omega\tau)\chi_{0}$, and as a consequence $\chi_{0}(\omega)'$ should dip and $\chi_{0}(\omega)''$ shows a maximum since the characteristic time $\tau$ of a ferromagnet diverges at the Curie temperature. However, $\chi_{0}(\omega)'$ in Fig. 1(b) does not dip but peaks at $\sim$ 37 K and $\sim$ 60 K, indicating that the nonlinear component
is becoming significant at the peak temperature ($T_{p}$) as in ferromagnets, spin-glass and cluster-glass materials \cite{Shirane,Koyano}.
In addition, the nonlinear susceptibilities $\chi_{2}'$ and $\chi_{2}''$ peak at the same temperature.
Therefore, the two maxima in $\chi_{0}'$ and $\chi_{0}''$ are related to ferromagnetic-paramagnetic phase transitions and there are two different magnetic phases in the sample. The behavior we observed is similar to previous reports on Fe$_{x}$TiS$_{2}$ single crystals \cite{Koyano} and manganite perovskite La$_{1-x}$Ca$_{x}$MnO$_{3}$ single crystals \cite{Markovich}, exhibiting spin-glass or cluster-glass and magnetic phase separation, respectively.
Since all the samples show two peaks, we define the peak temperature far below $T_{c}$ as $T_{p}$($K_{c}$) and that near $T_{c}$ as $T_{p}$($K_{u}$) (reasons given below).
\begin{figure}
\includegraphics[width=8cm]{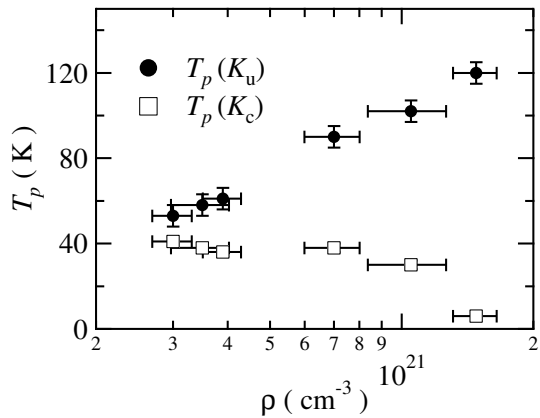}%%
\caption{Plots of $T$$_{p}$($K$$_{u}$) and $T$$_{p}$($K$$_{c}$) measured in $h =$ 6 Oe at $\omega =$ 1 Hz vs hole concentration $p$ at room temperature.}
\end{figure}

Comparing the temperature-dependent magnetic anisotropy with the ac susceptibility, we notice that the temperature where crossover of the magnetic easy axis occurs is close to $T_{p}$($K_{c}$) in the linear susceptibility so that $\left\langle 100 \right\rangle$ cubic magnetocrystalline anisotropy and [110] uniaxial anisotropy are likely to be associated with the two magnetic phases. In this regard, we assume that one of the magnetic phases with $T_{p}$($K_{c}$) has $\left\langle 100 \right\rangle$ cubic magnetocrystalline anisotropy while the other with $T_{p}$($K_{u}$) has [110] uniaxial anisotropy. This assumption is compatible with the fact that $\left\langle 100 \right\rangle$ cubic magnetocrystalline anisotropy is dominant at $T \lesssim$ 25 K whereas [110] uniaxial anisotropy is enhanced with increasing temperature. Also, the experimental results of magnetization reversal processes in (Ga,Mn)As, which indicate a significant contribution of [110] uniaxial anisotropy above 40 K, support the assumption\cite{Hamaya2}. Hereafter, we term phases with cubic anisotropy and uniaxial anisotropy the cubic phase and uniaxial phase, respectively.

Our recent experiment \cite{Hamaya4} and theoretical prediction reported by Dietl {\it et al.} \cite{Dietl2} indicated a strong correlation between magnetic anisotropy and hole concentration in (Ga,Mn)As. In order to further examine the correlation between magnetic transition temperatures  and hole concentration, we plot $T_{p}$($K_{c}$) and $T_{p}$($K_{u}$) obtained  from $\chi_{0}''(T)$ curves in $h =$ 6 Oe at $\omega =$ 1 Hz as a function of hole concentration in Fig. 2. Note that the data in Fig. 2 were taken from a sample before and after annealing with a fixed as well as a different Mn content. $T_{p}$($K_{u}$) clearly increases with increasing hole concentration, while $T_{p}$($K_{c}$) is reduced, in particular, in the high hole-concentration regime.

It has been demonstrated that magnetic anisotropy and magnetization reversal processes are qualitatively interpreted with the following equation \cite{Hamaya2,Hamaya4}, i.e., $E$ $=$ $K_{u}$sin$^{2}$($\varphi$ $-$ 45$^\circ$) $+$ ($K_{c}$/4) sin$^{2}$2$\varphi$ $-$ $MH$ cos($\varphi$ $-$ $\theta$), where {\it K$_{u}$} and {\it K$_{c}$} are the in-plane uniaxial and cubic anisotropy constants, {\it M} is the magnetization, {\it H} is the field strength, and $\varphi$ is the direction of magnetization with respect to [100] in the film plane. Assuming that the hard axis $M-H$ curves are based on coherent rotation \cite{Welp,Hamaya4}, we can estimate {\it K$_{u}$} and {\it K$_{c}$} using conditions $\partial${\it E}/$\partial$$\varphi$ $=$ 0, $\partial$$^2${\it E}/$\partial$$\varphi$$^2$ $>$ 0, and $\theta =$ 135$^{\circ}$ ([1$\overline{1}$0]) \cite{Hamaya4}.
Figure 3 shows both numerically calculated $K$$_{u}$/$K$$_{c}$ at 4 K and $T$$_{p}$($K$$_{u}$)/$T$$_{p}$($K$$_{c}$) obtained from the data in Fig. 2 as a function of hole concentration.
It should be noted that $K$$_{u}$/$K$$_{c}$ and $T$$_{p}$($K$$_{u}$)/$T$$_{p}$($K$$_{c}$) show a similar dependence on hole concentration with a marked increase in the high hole concentration regime. Therefore, this corroborates that $T$$_{p}$($K$$_{c}$) and $T$$_{p}$($K$$_{u}$) are attributed to cubic and [110] uniaxial anisotropy phases.
\begin{figure}[t]
\includegraphics[width=8.5cm]{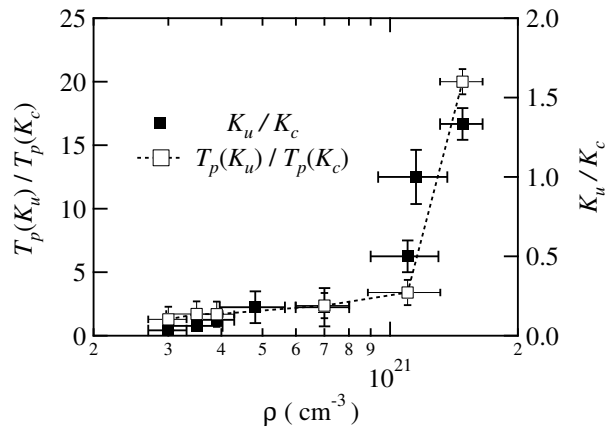}%%
\caption{Measured $T$$_{p}$($K$$_{u}$)/$T$$_{p}$($K$$_{c}$) and calculated $K$$_{u}$/$K$$_{c}$ as a function of hole concentration $p$.}
\end{figure}

Note that the feature at around 60 K has a strong frequency dependence while that at around 37 K has no frequency dependence (Fig. 1). A similar behavior has been reported in La$_{0.8}$Ca$_{0.2}$MnO$_{3}$ single crystals \cite{Markovich}, where a strong frequency dependence of the linear susceptibility was found due to the blocking process of spin-clusters in the La$_{0.8}$Ca$_{0.2}$MnO$_{3}$ matrix.
From the frequency dependence of $T$$_{p}$($K$$_{u}$), we deduce that magnetic phase with [110] uniaxial anisotropy, which we detected through ac susceptibility measurements, can be an assembly of (Ga,Mn)As magnetic clusters with a relatively high hole concentration and corresponding [110] uniaxial anisotropy in a (Ga,Mn)As matrix with low hole concentration and $T$$_{p}$($K$$_{c}$). Since the (Ga,Mn)As matrix has a low Curie temperature, the ferromagnetic clusters are embedded in nonmagnetic (Ga,Mn)As and the magnetic moments are blocked along  [110] in a temperature range of  $T$$_{p}$($K$$_{c}$)$<T<$$T$$_{p}$($K$$_{u}$) (Fig. 4(a)). When the temperature is raised to $T$ $>$ $T$$_{p}$($K$$_{u}$), the magnetic moments of the clusters become random due to melting of the blocked moments (Fig. 4(b)). At temperatures below $T$$_{p}$($K$$_{c}$), on the other hand, the cubic phase is now ordered ferromagnetically and competes with the ferromagnetism of the uniaxial cluster phase. If the [110] uniaxial anisotropy is relatively small, the magnetic easy axis of the whole epilayer should be governed by the cubic phase predominantly below $T$$_{p}$($K$$_{c}$) \cite{Hamaya2,Welp,Kato}.
\begin{figure}
\includegraphics[width=8cm]{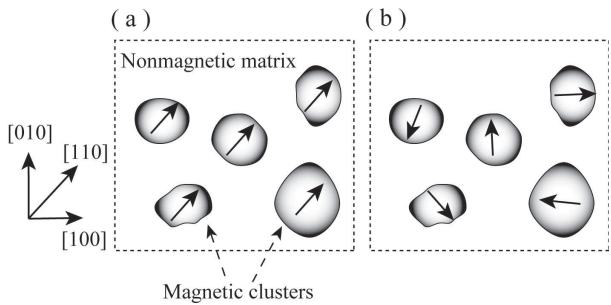}%%
\caption{Schematic model of ferromagnetic ordering in (Ga,Mn)As in ( a ) $T$$_{p}$($K$$_{c}$)  $<$ $T$ $<$ $T$$_{p}$($K$$_{u}$) and in ( b ) $T$ $>$ $T$$_{p}$($K$$_{u}$). Magnetic moments of clusters are aligned to [110] in $T$$_{p}$($K$$_{c}$)  $<$ $T$ $<$ $T$$_{p}$($K$$_{u}$) because the clusters have zinc-blende-type single phase.}
\end{figure}

In general, these magnetic clusters have a characteristic superparamagnetic frequency given by $f$ $=$ $f_{0}$ exp( $-$ $KV$/$kT$), where $f_{0}$ represents the specific frequency of attempts for a cluster to change its spin orientation, $K$ is the magnetic anisotropy constant, and $V$ is the activation volume of the magnetic cluster \cite{Markovich}. In the present case, Arrhenius plot $lnf$ vs $T$$_{p}$($K$$_{u}$)$^{-1}$ gives an activation energy of 0.73 eV to change the spin orientation.
Assuming that $K$ is 5.0 $\times$ 10$^{2}$ erg/cm$^{3}$ \cite{ref.}, a typical magnetic cluster is about 82 nm in diameter, which is much smaller than the magnetic domain size above $T_{c}$/2 observed in a previous report \cite{Welp}. However, the domain imaging technique averages magnetic structures over an area with 1$\mu$m in size so that our description is consistent with the observation of domain structures , provided that all clusters align crystallographically in the epilayer.

Now we discuss the process by which (Ga,Mn)As clusters are formed with [110] uniaxial anisotropy and the matrix is formed in (Ga,Mn)As epilayers. Mn interstitials generated during low-temperature growth processes are believed to diffuse toward substitutional Mn acceptors as a results of post-annealing due to local electric field \cite{Edmonds}. This diffusion induces the local accumulation of substitutional Mn in the epilayer. The accumulation has also been inferred from a recent report by Welp {\it et al.}, who proposed that [110] uniaxial anisotropy is due to preferential Mn incorporation induced by the presence of As dimers in every monolayer of the layer-by-layer grown epilayer \cite{Welp2}.
Taking these considerations and our experimental results into account, the diffusion process facilitates the formation of highly substituted (Ga,Mn)As clusters in (Ga,Mn)As epilayers, in particular, in the high hole-concentration regime. On the other hand, the (Ga,Mn)As matrix is likely to have relatively low ionized Mn content, resulting in $\left\langle 100 \right\rangle$ cubic anisotropy and a low Curie temperature. The description is also supported by our recent results, in which a significant enhancement in [110] uniaxial anisotropy is induced by low-temperature annealing in a sample with fixed Mn content \cite{Kato}.

In conclusion, we have found that (Ga,Mn)As epilayers have two different ferromagnetic phases through ac susceptibility and magnetization measurements. A comparison of ac susceptibility with numerical calculation of the magnetic anisotropy constants indicates that phase with a lower  ferromagnetic-paramagnetic transition temperature $T$$_{p}$($K$$_{c}$) has $\left\langle 100 \right\rangle$ cubic magnetocrystalline anisotropy while phase with a higher ferromagnetic-paramagnetic transition temperature $T$$_{p}$($K$$_{u}$) shows [110] uniaxial anisotropy. The frequency dependence of the ac susceptibility have revealed that $\left\langle 100 \right\rangle$ cubic anisotropy and [110] uniaxial anisotropy are attributed to a (Ga,Mn)As matrix and highly-substituted (Ga,Mn)As clusters, respectively. These results clearly demonstrate that a change in the magnetic easy axis from [100] to [110] with increasing temperature is a consequence of the marked decrease in the contribution of $\left\langle 100 \right\rangle$ cubic anisotropy at around $T$$_{p}$($K$$_{c}$).

%\section{CONCLUSION}
%\section{ACKNOWLEDGMENTS}
We acknowledge Prof. Hiro Munekata for providing (Ga,Mn)As samples used in this study. 
K. H. would like to thank Dr. A. Oiwa for his useful discussions and acknowledge the financial support given by the 21st Century COE Program, ``Nanomaterial Frontier Cultivation for Industrial Collaboration", of Ministry of Education, Culture, Sports, Science and Technology (MEXT), Japan. This work was partially supported by the Scientific Research in Priority Areas ``Semiconductor Nanospintronics", provided by MEXT.

%\begin{acknowledgments}
%\end{acknowledgments}

\vspace{5mm}
\noindent{REFERENCES}


\begin{thebibliography}{99}
\bibitem{ref.}{The magnetic anisotropy constant $K$ is chosen from our recent study \cite{Hamaya4} which has estimated the value using magnetization curves and magnetoenergetics calculations.}
\end{thebibliography}
\end{document}